# Linear and angular momenta of photons in the context of "which path" experiments of quantum mechanics

Masud Mansuripur

James C. Wyant College of Optical Sciences, The University of Arizona, Tucson

**Abstract**. In optical experiments involving a single photon that takes alternative paths through an optical system and ultimately interferes with itself (e.g., Young's double-slit experiment, Mach-Zehnder interferometer, Sagnac interferometer), there exist fundamental connections between the linear and angular momenta of the photon on the one hand, and the ability of an observer to determine the photon's path through the system on the other hand. This paper examines the arguments that relate the photon momenta (through the Heisenberg uncertainty principle) to the "which path" (German: welcher Weg) question at the heart of quantum mechanics. We show that the linear momenta imparted to apertures or mirrors, or the angular momenta picked up by strategically placed wave-plates in a system, could lead to an identification of the photon's path only at the expense of destroying the corresponding interference effects. We also describe a thought experiment involving the scattering of a circularly-polarized photon from a pair of small particles kept at a fixed distance from one another. The exchange of angular momentum between the photon and the scattering particle in this instance appears to provide the "which path" information that must, of necessity, wipe out the corresponding interference fringes, although the fringe-wipe-out mechanism does not seem to involve the uncertainty principle in any obvious way.

**1. Introduction**. In the historic Bohr-Einstein debates concerning the foundations of quantum mechanics,[1] the Young double-slit experiment of classical optics played a central role. In a nutshell, Einstein argued that the passage of a single photon could be attributed to one slit or the other, since the mechanical momentum picked up by the plate housing the slits contains the information needed to identify the path taken by the photon through the system. Bohr's counter argument was that acquiring the "which path" information by way of monitoring the plate's momentum would, in accordance with Heisenberg's uncertainty principle, perturb the position of the plate, thus wiping out the expected interference fringes.[2] In other words, acquisition of the "which path" information, while revealing the particle-like nature of the photon, would blur the anticipated interference fringes to the extent that all evidence for the photon's wave-like behavior will be lost.

The goal of the present paper is to examine a few variations on the theme of the Bohr-Einstein argument, and to explore the extent to which Heisenberg's uncertainty principle needs to be relied upon in the context of "which path" experiments. Feynman has emphasized that, in the absence of "which path" information, it is the (complex) probability amplitudes of an event that must be added together, whereas the existence of such information, even in principle, compels one to explain the outcome of an experiment by directly adding the probabilities associated with individual paths that could have led to a specific event.[3] Thus, wave-like behavior is exhibited when the probability amplitudes are required to be added together, whereas particle-like behavior emerges in situations where the probability of occurrence of a multi-path event turns out to be the sum of the probabilities of its individual paths. We will see in the next section how the availability of "which path" information (in the form of a photon's polarization state in the double-slit experiment) wipes out the interference fringes, and also how the "erasure" of this information causes the fringes to reappear.

A photon of frequency $\omega$ propagating in a given direction in free space, say, along the $z$-axis, is a wavepacket of energy $\hbar\omega$ and linear momentum $(\hbar\omega/c)\hat{\mathbf{z}}$ in the number state $|1\rangle$; here $\hbar$ is the reduced Planck constant and $c$ is the speed of light in vacuum.[4] In addition, a circularly-polarized photon carries angular momentum in the amount of $\pm\hbar\hat{z}$, with the plus or minus sign depending on the sense (or helicity) of its circular polarization. While deflection of a photon's trajectory at a slit, or its reflection from a conventional mirror, involves an exchange of linear momentum, its passage through a birefringent plate could entail an exchange of angular momentum as well.[5] We will discuss examples of systems where the angular momentum transferred to a wave-plate or to a scatterer can be relied upon to provide the desired "which path" information.

The organization of the paper is as follows. In Sec.2, we examine the canonical Young's double-slit experiment[6,7] in the special case when a single photon passes through the pair of slits and interferes with itself at the observation plane. Reconstructing Bohr's argument in his debates with Einstein, we show how a straightforward application of Heisenberg's uncertainty principle confirms the incompatibility of observing wave-like behavior by the photon (i.e., appearance of interference fringes) with particle-like behavior (namely, acquisition of information about the slit through which the photon has passed).[1] We proceed to extend the analysis to cases where a $\pi$ phase-shifter is placed in one of the slits, or when a birefringent window is placed in a slit, to shed additional light on the nature of Bohr's complementarity principle.

Section 3 is devoted to an analysis of a single photon passing through a Mach-Zehnder interferometer,[6,7] where the probability amplitudes associated with its passage through one arm or the other of the device combine to reveal the possibility of observing the effects of interference at the output ports of the interferometer. It will be seen once again that the acquisition of "which path" information (by monitoring the mechanical momenta picked up by the mirrors within the device) disturbs the optical path lengths just enough to eliminate the possibility of observing the anticipated interference. An alternative method of acquiring the "which path" information (based on placing a quarter-wave plate in each arm of the interferometer, then monitoring their angular momenta before and after the passage of the photon) will be shown to similarly ruin the observability of interference.

In the case of the Sagnac interferometer[8-10] examined in Sec.4, the mechanical momenta imparted to the mirrors by a passing photon (or any angular momentum picked up by the device as a whole) do not contain the desired "which path" information. However, a pair of properly oriented and strategically placed quarter-wave plates (QWPs) in the system can, in principle, pick up angular momenta in the amount of $\pm\hbar$ from their interaction with the passing photon. Once again, application of the uncertainty principle reveals the impossibility of acquiring "which path" information without disturbing the necessary conditions for observing the interference effects associated with the photon's wave-like behavior.

In Sec.5, we describe a thought experiment involving a right-circularly-polarized (RCP) photon being scattered from a pair of small particles that are kept at a fixed distance from each other. The existence of a nonzero probability for the scattered photon to become left-circularly-polarized (LCP) indicates that the difference $2\hbar$ between the angular momenta of the incident and scattered photons must be picked up by one of the two scatterers. The availability of the "which path" information in this instance does not appear to disturb the state of the scattered photon in ways that would prohibit the formation of interference fringes at the observation plane. While invoking the Heisenberg uncertainty principle in this case does not bring about the disturbances needed to prevent fringe formation, the entanglement of the state of the pair of particles with that of the scattered photon guarantees that interference will not occur and that, therefore, the core principle of wave-particle duality at the heart of quantum mechanics remains inviolate.

The paper closes in Sec.6 with a summary and a few concluding remarks. An Appendix provides a concise derivation of the Heisenberg uncertainty relation between a pair of non-commuting observables.

**2. Young's double-slit experiment.** Figure 1 shows a slight variation on the Young double-slit arrangement,[6,7] where the incident light is a single-photon wavepacket in the number state $|1\rangle$, having frequency $\omega$, $k$-vector $\boldsymbol{k} = (\omega/c)\hat{\boldsymbol{z}}$, and linear polarization $\hat{\boldsymbol{e}} = \hat{\boldsymbol{x}}$. (In terms of the vacuum wavelength $\lambda$, the magnitude of $\boldsymbol{k}$ may be written as $k = 2\pi/\lambda$.) The slits of width $w$ are carved into two separate plates, which are symmetrically positioned in the $xy$-plane such that the center-to-center distance between the slits along the $x$-axis is $d$. The splitting into two separate parts of the plate that houses the slits is intended to simplify the forthcoming analysis of momentum transfer from the photon to the plates for purposes of identifying the slit through which the photon has passed. Let the $E$-field amplitude of the light transmitted through the pair of slits be written as

$$E(x, z = 0^+) = E_\text{o}\text{rect}(x/w) * [\delta(x + \tfrac{1}{2}d) + \delta(x - \tfrac{1}{2}d)], \qquad (1)$$

where $\delta(x)$ is Dirac's delta-function, and the standard function $\text{rect}(x)$ equals 1.0 when $|x| < \tfrac{1}{2}$, zero when $|x| > \tfrac{1}{2}$, and $\tfrac{1}{2}$ when $|x| = \tfrac{1}{2}$; the asterisk represents the convolution operation, and the Fourier transform of $\text{rect}(x)$ is $\int_{-\infty}^{\infty} \text{rect}(x) e^{-\text{i}2\pi sx} \text{d}x = \sin(\pi s)/(\pi s) = \text{sinc}(s)$. Thus, the Fourier transform of the emergent $E$-field amplitude immediately after the slits is given by

$$\tilde{E}(\sigma_x) = \int_{-\infty}^{\infty} E(x) e^{-\text{i}2\pi \sigma_x x} \text{d}x = 2E_\text{o} w \, \text{sinc}(w\sigma_x) \cos(\pi d \sigma_x). \qquad (2)$$

The product $\lambda \sigma_x$ of the incident wavelength $\lambda$ and the Fourier variable $\sigma_x$ appearing in Eq.(2) represents $\sin \theta$ for an emergent plane-wave (corresponding to a geometric-optical ray) at an angle $\theta$ relative to the $z$-axis; see Fig.1. Thus, in the observation plane in the far field, the $\pm 1^\text{st}$ peaks of the fringe pattern appear at $\sin \theta = \pm \lambda/d$. To ensure that these fringes do not get blurry (or washed out), the positions of the two plates that house the slits must each have an uncertainty $\Delta x$ along the $x$-axis much less than $d$, say, $\Delta x = \alpha d$ with $\alpha \lesssim 0.1$. The visibility of the fringes that are further away from the center of the observation plane requires progressively smaller values of $\alpha$.

A photon arriving at the center of a first bright fringe on one side or the other of the central fringe will have the following $x$-component of momentum:

$$(\hbar \omega/c) \sin \theta = \pm(2\pi\hbar/\lambda)(\lambda/d) = \pm 2\pi\hbar/d. \qquad (3)$$

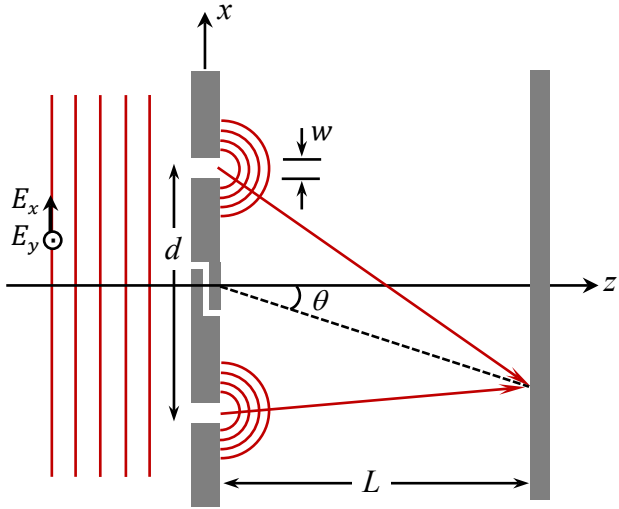

**Fig.1**. A slight variation on Young's double-slit experiment, where the plate containing the slits (width $= w$, separation $= d$) is split in the middle, so that, in the wake of a single photon's passage through either slit, the $x$-component of momentum transferred to the corresponding half-plate can be monitored.

Thus, if the photon is assumed to have passed through one of the slits, the plate containing that slit must have acquired a backlash momentum along the $x$-axis with an absolute value of $2\pi\hbar/d$. However, Heisenberg's uncertainty principle ($\Delta x \Delta p_x \geq \tfrac{1}{2}\hbar$) constrains the initial $x$-momentum of each plate to $\Delta p_x \gtrsim \hbar/(2\alpha d)$, making it impossible to determine which plate has picked up the $2\pi\hbar/d$ momentum of the transmitted photon. This is how complementarity in quantum mechanics ensures, through the uncertainty principle, that acquisition of "which path" information in the double-slit experiment is incompatible with the existence of interference fringes.[†]

Suppose now that a $\pi$ phase-shifter — not a birefringent plate, but simply an isotropic glass window that imparts a $\pi$ phase-shift to the incident photon — is placed inside one of the slits. The edges of this phase-shifting window are in contact with the interior walls of the slit and, therefore, any momentum exchange between the photon and the plates can be associated with the interface between the slit and the dielectric material that fills the slit and constitutes the phase-shifting window. As before, the photon's state of polarization is fixed (but otherwise arbitrary), and the effect of the phase-shifter is to make the emergent $E$-field amplitude an odd function of $x$, namely,

$$E(x, z = 0^+) = E_0 \text{rect}(x/w) * [\delta(x + \tfrac{1}{2}d) - \delta(x - \tfrac{1}{2}d)]. \tag{4}$$

The Fourier transform of the above distribution is given by

$$\tilde{E}(\sigma_x) = \mathrm{i} 2 E_0 w \, \text{sinc}(w\sigma_x) \sin(\pi d \sigma_x). \tag{5}$$

Thus, the linear momentum distribution emerging from the pair of slits differs from that in the absence of the phase-shifter. Consequently, the fringes appearing at the observation plane resemble those formed in the absence of the phase-shifter, albeit with a vertical shift by one-half of one fringe along the $x$-axis, which results in a dark fringe appearing at the center of the observation plane. Aside from this exchange of positions between bright and dark fringes, no differences exist between the interference patterns formed with and without a $\pi$ phase-shifter residing in one of the slits. The preceding arguments with regard to the "which path" information, momentum exchange between the incident photon and the plates, and the restrictions imposed by the uncertainty principle continue to hold. It is worth emphasizing, however, that the vertical momentum distribution of the emergent photon in consequence of its passage through the pair of slits has changed from that given by Eq.(2) in the absence of the glass window to that in Eq.(5) when the $\pi$ phase-shifter is present.

As an aside, let us point out that the slits may be considered as two optical waveguides and that, according to classical electrodynamics, an incident plane-wave launches guided modes inside each slit.[11] Each guided mode is a superposition of plane-waves having vertical $k$-vector projections (and therefore vertical momenta) inside each waveguide. When the guided modes emerge from the slits, the various plane-waves emerging from the two slits will have different phases relative to each other — due to the separation distance between the slits. Some of these plane-waves interfere constructively and produce bright fringes at the observation screen, while others interfere destructively and produce the dark fringes. The momentum exchange between the light and the plates that house the slits thus occurs as a result of two physical phenomena: (i) excitation of guided modes within individual slits, and (ii) interference between the two emergent modes on the exit side of the slits.

---

[†] If the plate containing the slits is not split in the middle, the monitor of its $x$-momentum would have to be sensitive enough to tell which slit the photon has passed through. In this case, the resulting uncertainty $\Delta x$ in the position of the plate exceeds the fringe spacing $L\lambda/d$ at the observation plane, so that, once again, the fringes will be washed out.[2]

As a final example, consider replacing the isotropic $\pi$ phase-shifter with a birefringent window that imparts a $\pi$ phase shift to the $E_y$ component of the incident polarization, but a net zero phase shift to the $E_x$ component. This birefringent window (placed, as before, in only one of the slits) thus acts as a half-wave plate that causes the emergent photon's $E_x(x, z = 0^+)$ to resemble that in Eq.(1), while the emergent $E_y(x, z = 0^+)$ acquires the general from of Eq.(4).[12] If the incident photon is assumed to be linearly polarized at 45° to the $x$-axis (i.e., a superposition of equal amounts of $x$ and $y$ polarization), the resulting $E$-field profile at the observation plane will consist of a set of $x$-polarized fringes that have a bright central fringe, superposed on a set of $y$-polarized fringes whose central fringe is dark. Considering that the $x$-polarized bright fringes cover the $y$-polarized dark fringes and vice-versa, the observed intensity distribution at the observation plane lacks any bright and dark fringes, thus conveying the impression that the emergent light from the pair of slits is incapable of producing interference fringes. This, of course, is consistent with the fundamental dictum of quantum electrodynamics that requires the addition of probability amplitudes for a photon arriving at a given point in the observation plane after taking both alternative paths through the two slits provided that, in principle, the two paths are indistinguishable.[3] Considering that an incident photon's polarization changes from +45° to −45° (relative to the $x$-axis) if it passes through the slit that houses the half-wave plate, this subsequently detectable "which path" information that is carried by the photon in its own polarization state is sufficient to eliminate the possibility of optical interference.

Note in the preceding example that the polarization distribution along the $x$-axis at the observation plane is not uniform; that is, the $E$-field continuously varies from a linearly-polarized state along the $x$-axis to a state that is elliptical, then circular, then elliptical again, then linear along the $y$-axis, and so forth; this pattern repeats as one moves up or down along the $x$-axis. In other words, the interference effects are "hidden" inside the polarization state of the photon that arrives at the observation plane, even though the probability of detecting the photon as a function of position along the $x$-axis does not exhibit any interference effects. In the context of photodetection probability, one can recover the interference fringes by erasing the "which path" information, say, by placing a sheet polarizer before the observation plane that prevents either the $x$-polarized or the $y$-polarized photons from reaching the photodetector.[12]

The above examples reveal the intimate connection between the existence of interference fringes (wave-like behavior) and the availability of "which path" information (particle-like behavior) for single photons in a double-slit experiment. While an exchange of linear momentum between the incident photon and the pair of slits takes place under all circumstances, the exact amount of the transferred momentum depends on the presence or absence of phase-shifting windows within each slit and, in the case of birefringent windows, on the polarization state of the incident photon. With regard to the momentum exchange process, the crucial point to keep in mind is not just the modes excited inside each slit (treated as a waveguide), but also interference between the guided modes upon emergence from the two slits.

**3. The Mach-Zehnder interferometer**. A wave-packet containing a single photon of frequency $\omega$ in the number state $|1\rangle$ enters a Mach-Zehnder interferometer,[6,7] as shown in Fig.2. If the optical path-length difference between the two arms of the device is properly adjusted, either constructive or destructive interference will take place at the second beam-splitter, and the photon consistently emerges from one or the other exit channel of the interferometer. In what follows, we shall explain why it is impossible to monitor the mechanical momentum acquired by one of the mirrors (say, the retro-reflecting mirror) afterward in order to determine the path taken by the photon.

It is well known that the incident photon has energy $\hbar\omega$ and linear momentum $\hbar\omega/c$ along its propagation direction in free space. If the retro-reflecting mirror's momentum $\boldsymbol{p}_M$ (after the passage of the photon through the system) is to convey information about the path of the photon, then the uncertainty concerning the value of $p_M$ before the photon's arrival must be far less than the photon momentum; that is, $\Delta p_M \ll \hbar\omega/c$. The uncertainty $\Delta x_M$ in the mirror's position then, according to Heisenberg's principle, must satisfy the relation $\Delta x_M \Delta p_M \geq \tfrac{1}{2}\hbar$. Consequently, $\Delta x_M \gg c/2\omega = \lambda/4\pi$. This level of uncertainty in the retroreflector's position is sufficient to destroy the necessary conditions for constructive or destructive interference. As a result, the photon will randomly emerge from one or the other of the output channels, thus preempting the possibility of observing optical interference, the characteristic signature of the photon's wave-like behavior.

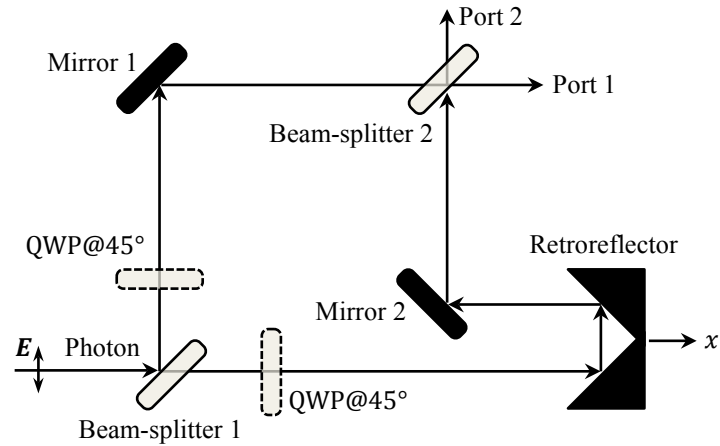

**Fig.2**. A linearly-polarized wave-packet containing a single photon of energy $\hbar\omega$ enters a Mach-Zehnder interferometer, bounces off the various reflectors, and eventually exits through port 1 or port 2, depending on the optical path-length difference between the two arms of the device. Shifting the retroreflector along the $x$-axis enables one to adjust the path-length along the lower arm of the interferometer. In the absence of the quarter-wave plates, the EM field remains linearly polarized throughout the system. Inserting a QWP into each arm with their fast and slow axes oriented at 45° to the direction of the incident $E$-field, converts the polarization state of the photon from linear to circular. Interference at beam-splitter 2 continues to be constructive or destructive (depending on the position of the retroreflector along the $x$-axis) provided that the sense of circular polarization (i.e., right or left) is the same in both arms.

Note that a possible way to ensure the position accuracy of a given mirror is to endow it with a large mass so that, for example, an optical measurement of its position will not disturb its position. A large mass, however, will introduce a large uncertainty in the initial momentum of the mirror, so that the post-passage measurement of the mirror's momentum in the Mach-Zehnder system will not yield any useful information about the path taken by the photon.

We mention in passing that, to ensure a substantial overlap between the wavepackets that arrive at the second splitter, the path-length difference between the two arms of the interferometer must be well below the length of the photon's wavepacket along its propagation path. For an estimate of the wavepacket's length, one may use either of the uncertainty relations $\Delta\mathcal{E}\Delta t \gtrsim \tfrac{1}{2}\hbar$ or $\Delta x \Delta p_x \gtrsim \tfrac{1}{2}\hbar$, and assume a frequency linewidth $\Delta\omega$ around, say, one percent of the photon's center frequency. For instance, if the center frequency is taken to be $\omega_o = 3.77 \times 10^{15}$ rad/sec (corresponding to a photon wavelength of $\lambda_o = 0.5$ μm), we will have $\Delta t \cong 30$ fs or $\Delta x \cong 9$ μm.

As an alternative method of acquiring "which path" information about the putative path taken by the incident photon through the Mach-Zehnder interferometer, consider inserting a quarter-wave plate (QWP) in each arm of the device at 45° to the direction of incident (linear) polarization, so that both packets become circularly polarized — both right-circular or both left-circular.[‡] This does not affect the overall behavior of the system, as the interference phenomenon continues to work in the same way for circular as it does for linear polarization. However, in the case of a single photon passing through the system, the "which path" question may be answered by examining the angular momentum picked up by one or the other QWP. A version of Heisenberg's uncertainty principle (described below) then tells us that the need to measure the spin angular momentum $\hbar$ picked up by either of the QWPs requires an uncertainty in the initial orientation of the wave-plate that would bring its output to an unspecified state of polarization (i.e., linear, circular, or elliptical). Once again, this lack of certainty in the initial orientation of the QWPs destroys the possibility of observing the effects of interference at the output ports of the interferometer.

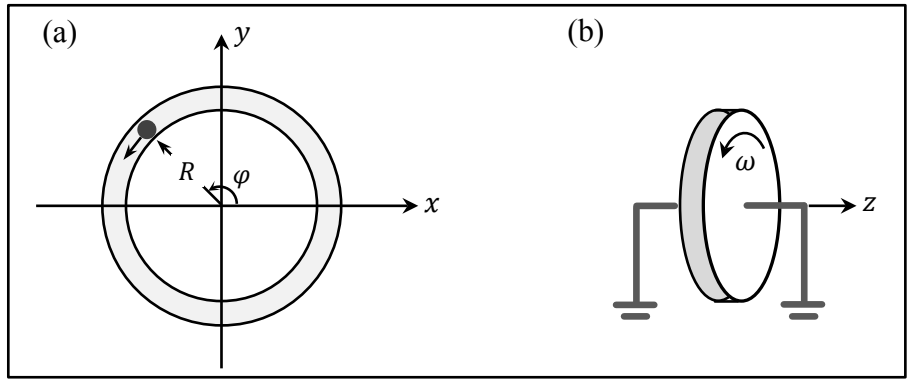

**Fig.3**. (a) A small particle of mass $m$ moves with constant angular velocity $\omega\hat{\mathbf{z}}$ around a circular track of radius $R$. The particle's linear and angular momenta are $\boldsymbol{p} = mR\omega\hat{\boldsymbol{\varphi}}$ and $\boldsymbol{L} = mR^2\omega\hat{\mathbf{z}}$, respectively. (b) A solid disk having moment of inertia $I$ rotates around the $z$-axis with angular velocity $\omega$, angular momentum $\boldsymbol{L} = I\omega\hat{\mathbf{z}}$ and kinetic energy $\mathcal{E} = \tfrac{1}{2}I\omega^2$. In both cases (a) and (b), the uncertainty $\Delta L_z$ in the angular momentum and the uncertainty $\Delta\varphi$ in the angular position are related via $(\Delta L_z)(\Delta\varphi) \gtrsim \tfrac{1}{2}\hbar$.

While there is no known principle that would relate a measure of uncertainty $\Delta L$ in the angular momentum to that of the orientation angle $\Delta\varphi$ of a QWP, an argument can be advanced to infer an approximate relation between $\Delta L$ and $\Delta\varphi$ from the well-known position-momentum uncertainty relation $(\Delta x)(\Delta p_x) \geq \tfrac{1}{2}\hbar$, as follows. With reference to Fig.3(a), let a small particle of mass $m$ moving around a circular track of radius $R$ within the $xy$-plane have angular velocity $\omega\hat{\mathbf{z}}$, linear momentum $\boldsymbol{p} = mR\omega\hat{\boldsymbol{\varphi}}$, and angular momentum $\boldsymbol{L} = mR^2\omega\hat{\mathbf{z}}$. Given a relatively small amount of uncertainty $\Delta\varphi$ in the particle's angular position, Heisenberg's uncertainty principle dictates that the product $(\Delta p_\varphi)(R\Delta\varphi)$ cannot be less than $\tfrac{1}{2}\hbar$, which is equivalently written as $(\Delta L_z)(\Delta\varphi) \gtrsim \tfrac{1}{2}\hbar$.

Alternatively, one can invoke the time-energy uncertainty relation $\Delta\mathcal{E}\Delta t \geq \tfrac{1}{2}\hbar$ in conjunction with the rotating disk depicted in Fig.3(b) to arrive at a similar relation between $\Delta L_z$ and $\Delta\varphi$ of the disk. This is because the rotational kinetic energy of the disk $\mathcal{E} = \tfrac{1}{2}I\omega^2 = L_z^2/2I$ yields $\Delta\mathcal{E} \cong L_z\Delta L_z/I$. Considering that $L_z\Delta t = I\omega\Delta t \cong I\Delta\varphi$, we arrive at $\Delta\mathcal{E}\Delta t \cong (\Delta L_z)(\Delta\varphi) \gtrsim \tfrac{1}{2}\hbar$.

Now, if the initial angular momentum of a QWP is known to an accuracy better than $\hbar$ to be close to zero, the uncertainty in its initial orientation must be $\Delta\varphi > \tfrac{1}{2}$ radian, which is sufficient to render the polarization state of the emergent photon highly uncertain. Consequently, if measurements performed on a QWP before and after the putative passage of a photon could reveal

---

[‡] A mirror or a beam-splitter changes the sense of circular polarization (from right to left and vice-versa) upon reflection. However, all will be good in the end, since there are two splitters as well as an odd number of mirrors in each arm of the Mach-Zehnder interferometer.

(with reasonable certainty) a change of $\hbar$ in the plate's angular momentum, the "which path" information thus attained would not contradict the fundamental tenets of quantum mechanics, since, under the circumstances, no interference would have occurred.

**4. The Sagnac interferometer**. In the case of a Sagnac interferometer, such as that depicted in Fig.4, a single photon entering the device at the beam-splitter $S$ follows both a clockwise and a counterclockwise path around the triangular loop, then emerges at the observation plane if there happens to be a $\pi$ phase difference between the two paths (brought about by the rotation of the plane of the interferometer).[8-10] In contrast to the case of the Mach-Zehnder interferometer, one cannot rely on the linear momenta picked up by the individual mirrors or by the beam-splitter of the Sagnac device to acquire the "which path" information. Nor can one glean such information from the angular momentum picked up by the Sagnac device as a whole, simply because such an angular momentum (with respect to any arbitrary point in space) equals the difference between the angular momentum of the incident photon and that of the emergent photon, which are path-independent.

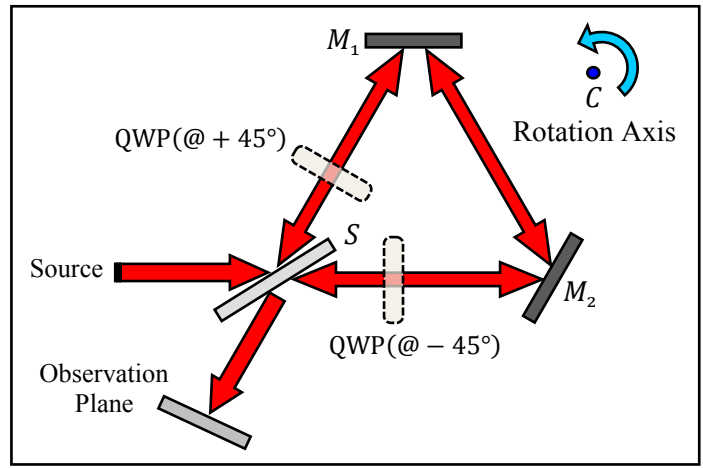

**Fig.4**. In a Sagnac interferometer, the light from the source is split at a 50/50 beam-splitter $S$ to travel in opposite directions around a loop formed by the splitter $S$ and the mirrors $M_1$ and $M_2$. Upon returning to the splitter, one beam is reflected and the other one transmitted at $S$, so that a superposition of the two arrives at the observation plane. With the interferometer standing still, the counter-propagating beams around the $SM_1M_2$ loop remain in phase; however, one beam suffers two reflections at the splitter while the other one gets transmitted twice; the net relative phase between the two beams will then be 180° and, therefore, no light reaches the observation plane. In contrast, when the interferometer rotates around an axis (say, one that crosses the $SM_1M_2$ plane at $C$), the counter-propagating beams acquire a relative phase that allows a fraction of the circulating light to reach the observation plane. The quarter-wave plates (QWPs) are not needed for the usual operation of the device, but are used here as a means of acquiring the "which path" information.

One way to obtain the "which path" information from a Sagnac interferometer is to place a pair of QWPs on both sides of the splitter $S$, as shown in Fig.4, then rely on the exchange of spin angular momentum between the photon and these wave-plates. In the triangular Sagnac device depicted in Fig.4, which consists of a single 50/50 splitter $S$ and two ideal mirrors $M_1$ and $M_2$, let a first QWP, oriented at $+45°$ relative to the direction of incident (linear) polarization, reside between the splitter and the first mirror, and a second QWP, oriented at $-45°$ relative to the direction of incident polarization, reside between the splitter and the second mirror. Upon reflection from the two mirrors, the state of circular polarization of the photon will remain unchanged. Thus, under normal circumstances, the photon's polarization state, in consequence of passing through both QWPs in either of its counter-propagating paths, will have rotated by 90°, which leaves the overall behavior of the device intact. The "which path" information may now be obtained by monitoring the angular momenta of the QWPs, since both will have acquired an angular momentum of $\hbar$ and will be rotating, albeit in opposite directions, after the passage of the photon. The sense of rotation of each plate, of course, will depend on whether the photon has taken a clockwise or a counterclockwise path through the Sagnac loop. As we argued in the preceding section in the case of the Mach-Zehnder interferometer, the uncertainty principle applied to $\Delta L$ and $\Delta \varphi$ of the QWPs guarantees that all interference effects will be lost if the specific path taken by the photon could be identified.

**5. Will there be interference when a photon scatters from a pair of small particles?** Shown in Fig.5 is a pair of small spherical particles separated from each other by a fixed distance $d$ along the $x$-axis. Also shown is a single right-circularly-polarized (RCP) photon that propagates along the $z$-axis and gets scattered by the pair of particles. The particles are homogeneous and isotropic, having polarizability $\alpha$, and the scattering amplitudes for RCP (+) and LCP (−) photons arriving at a point $x$ within an observation plane located a large distance $z_0$ from the particles are given by[13]

$$E_{\text{out}}^{(\pm)} = \frac{\alpha k_0^2 (\sin\theta \pm 1)}{8\pi\varepsilon_0} \left( \frac{e^{ik_0 r_1}}{r_1} + \frac{e^{ik_0 r_2}}{r_2} \right) E_{\text{in}}^{(+)}. \tag{6}$$

Here, $k_0 = \omega/c$ is the magnitude of the photon's $k$-vector in vacuum, and $\varepsilon_0$ is the permittivity of free space. The distances $r_1$ and $r_2$ from the particles to the observation point differ by $\sim d \cos\theta$, which is small enough to be neglected where $r_1$ and $r_2$ appearing in the denominators in Eq.(6) are concerned, but needs to be taken into account in the phase-factors $e^{ik_0 r_1}$ and $e^{ik_0 r_2}$. Thus, the phase difference associated with the two paths to the observation point $(x, z_0)$ is $\sim e^{ik_0 d \cos\theta}$. We conclude that, for an incident RCP photon, the probability amplitude for an RCP or LCP photon to arrive at the observation point should be

$$E_{\text{out}}^{(\pm)}/E_{\text{in}}^{(+)} \cong \frac{\alpha k_0^2 e^{ik_0 r_0}}{4\pi\varepsilon_0 r_0} (\sin\theta \pm 1) \cos(\pi dx/\lambda_0 r_0). \tag{7}$$

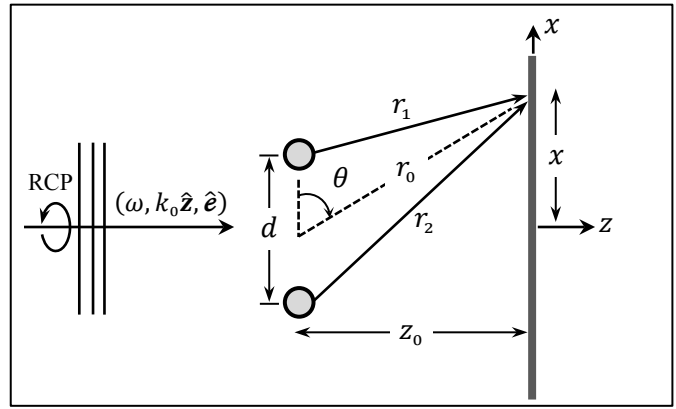

**Fig. 5**. Two small spherical particles at a fixed separation distance $d$ are suspended, say, in outer space, where gravitational as well as other forces are absent. A single RCP photon propagating along the $z$-axis is scattered from the particles and subsequently detected at a point $x$ in a distant observation plane. The uncertainties $\Delta x, \Delta y, \Delta z$ about the positions of the particles are sufficiently small, and the corresponding momentum uncertainties $\Delta \boldsymbol{p}$ sufficiently large, so that any measurement of the particles' linear momenta after the scattering event will not be able to identify the particle that scattered the photon. However, the angular momentum imparted by the scattered photon to either particle may be measureable, in which case the "which path" information acquired by monitoring the particles' angular momenta should prevent the formation of interference fringes at the observation plane.

At the center of the observation plane (i.e., in the vicinity of the z-axis), $\theta$ is close to 90°, which means that there is scant probability of receiving scattered LCP photons in this region. However, as one moves up or down along the x-axis, $\sin\theta$ is reduced, giving LCP photons a reasonable chance of being detected in regions where $|x|$ is large. Note that the conversion of an incident RCP photon into a scattered LCP photon entails a change of angular momentum, which should be accompanied by a transfer of an equal but opposite angular momentum to the scatterer.

An interesting question arises if the angular momenta of the small spherical particles could be monitored before and after the scattering. There does not appear to be any relevant uncertainty about the initial angular momenta of the particles that could prevent their accurate measurement following a scattering event. Consequently, the "which path" information gained by monitoring the angular momenta of the scatterers should, in principle, prevent the possibility of interference fringe formation at the observation plane.[2,14-16]

Initially, the particle-pair is in its ground-state $|s_0\rangle$, while the incident photon, prepared as a single-mode wavepacket $(\omega, k_0\hat{z}, \hat{e})$, occupies the number state $|1\rangle$; here, the complex unit-vector $\hat{e} = e' + ie''$ represents the photon's right-circular polarization state. Upon a scattering event in which the photon's polarization changes from RCP to LCP, the photon will be in a superposition state involving the single-mode wavepackets $(\omega, \mathbf{k}_1, \hat{e}^*)$ and $(\omega, \mathbf{k}_2, \hat{e}^*)$, with its joint number-state being $(|1\rangle_1|0\rangle_2 + |0\rangle_1|1\rangle_2)/\sqrt{2}$. However, the particle pair will now be in a new state, either $|s_1\rangle$ or $|s_2\rangle$, depending on whether it was the first or the second particle that caused the scattering—and, consequently, absorbed the change in the photon's spin angular momentum. Thus, the entangled state of the system comprising the scattered photon and the pair of scattering particles is found to be

$$|\psi\rangle = (|s_1\rangle|1\rangle_1|0\rangle_2 + |s_2\rangle|0\rangle_1|1\rangle_2)/\sqrt{2}. \tag{8}$$

Considering that $\langle s_1|s_2\rangle = 0$, the photodetection probability[4] for the LCP photon at the observation point $\mathbf{r}$ now lacks the cross-term proportional to $\cos[(\mathbf{k}_1 - \mathbf{k}_2) \cdot \mathbf{r}]$ that would be present in the absence of entanglement, hence the disappearance of interference fringes from the observation plane. Experimental evidence involving the scattering of polarized photons from a pair of trapped mercury ions in an ion trap can be cited in support of the above conclusion.[17]

**6. Concluding remarks**. Niels Bohr's complementarity principle holds that objects have certain pairs of complementary properties which cannot both be observed or measured simultaneously. The wave-particle duality of single photons is an example of such behavior whence setting up an experiment to reveal the particle-like behavior excludes the possibility of observing wave-like behavior and vice-versa; yet, understanding both experiments is essential if one hopes to fully grasp the nature of the photon.

This paper has examined three systems (namely, Young's double-slit experiment, the Mach-Zehnder interferometer, and the Sagnac interferometer) whose traditional mode of operation relies on the wave-nature of light as understood in classical optics, yet they are perfectly suited to operate in the presence of single photons [i.e., single-mode wavepackets $(\omega, \mathbf{k}, \hat{e})$ in the number state $|1\rangle$] and, upon repeated observations, reveal the wave-like behavior of such photons. In each instance, when attempts were made to observe a particle-like behavior (i.e., identify the path of the photon through the system), the operation of the instrument was shown to be substantially disturbed, to the point that the wave-like behavior could no longer be discerned. While it is the Heisenberg uncertainty principle that is usually called upon to explain the nature of the disturbance, the thought experiment presented in Sec.5 is believed to exhibit the essence of wave-particle duality without reliance on the uncertainty principle. The entanglement of the state of the scattered photon with that of the scattering particle pair appears to be all that is needed to destroy the interference fringes at the observation plane without the need to invoke the uncertainty principle.

# Appendix
# The Uncertainty Principle

Consider two Hermitian operators $\hat{A}$ and $\hat{B}$, whose commutator is given by $[\hat{A}, \hat{B}] = i\hat{C}$. Let the system at a given moment in time be in a state $|\psi\rangle$. We define the average, or expected value, of the observable $\hat{A}$ as well as its standard deviation $\delta_A$ as follows:

$$\langle A \rangle = \langle \psi | \hat{A} | \psi \rangle, \tag{A1}$$

$$\delta_A^2 = \langle \psi | (\hat{A} - \langle A \rangle)^2 | \psi \rangle = \langle A^2 \rangle - \langle A \rangle^2 = \langle \psi | \hat{A} \hat{A} | \psi \rangle - \langle A \rangle^2. \tag{A2}$$

Clearly, $\delta_A^2$ is a positive real number, because $|\psi\rangle$ can be expanded into a sum over the eigenstates of $\hat{A}$, namely, $|\psi\rangle = \sum_n c_n |\varphi_n\rangle$, in which case,

$$\langle A \rangle = \left(\sum_{n'} c_{n'}^* \langle \varphi_{n'}|\right) \hat{A} \left(\sum_n c_n |\varphi_n\rangle\right) = \left(\sum_{n'} c_{n'}^* \langle \varphi_{n'}|\right)\left(\sum_n c_n a_n |\varphi_n\rangle\right) = \sum_n a_n |c_n|^2, \tag{A3}$$

$$\delta_A^2 = \sum_n (a_n - \langle A \rangle)^2 |c_n|^2. \tag{A4}$$

Similar expressions may be written for the average $\langle B \rangle$ and the standard deviation $\delta_B$ of the observable $\hat{B}$ when the state of the system is $|\psi\rangle$. We now define the operators $\hat{A}_1 = \hat{A} - \langle A \rangle$ and $\hat{B}_1 = \hat{B} - \langle B \rangle$, whose average values vanish, but continue to have the same standard deviations, $\delta_A$ and $\delta_B$, and also the same commutator $[\hat{A}_1, \hat{B}_1] = i\hat{C}$ as $\hat{A}$ and $\hat{B}$. Let the operator $\hat{A}_1 - i\lambda\hat{B}_1$, where $\lambda$ is a real-valued parameter, act on $|\psi\rangle$. Since the norm of the resulting vector is positive, we will have

$$\langle \psi | (\hat{A}_1 + i\lambda\hat{B}_1)(\hat{A}_1 - i\lambda\hat{B}_1) | \psi \rangle = \delta_A^2 + \lambda^2 \delta_B^2 - i\lambda \langle \psi | [\hat{A}_1, \hat{B}_1] | \psi \rangle = \lambda^2 \delta_B^2 + \lambda \langle C \rangle + \delta_A^2 \geq 0. \tag{A5}$$

Note that $\hat{C}$, itself a Hermitian operator, must have a real-valued average $\langle C \rangle$. The second-order polynomial in $\lambda$ appearing on the right-hand-side of Eq.(A5) can be non-negative, irrespective of the value of $\lambda$, only if $\langle C \rangle^2 - 4\delta_A^2 \delta_B^2 \leq 0$. Thus follows the uncertainty relation $\delta_A \delta_B \geq \frac{1}{2}|\langle C \rangle|$. In the special case of the position and momentum operators, $\hat{x}$ and $\hat{p}_x$, where $[\hat{x}, \hat{p}_x] = i\hbar$, we obtain the famous Heisenberg uncertainty relation $\delta_x \delta_p \geq \frac{1}{2}\hbar$.

The other well-known uncertainty relation, namely, $\delta_\mathcal{E} \delta_t \geq \frac{1}{2}\hbar$, which relates the uncertainty in energy $\mathcal{E}$ to the uncertainty in time $t$, is not so easy to derive, because, in quantum mechanics, no operator is associated with time. One way to see the physical reasoning behind this uncertainty relation is to consider a system with a *time-independent* Hamiltonian $\hat{H}$ and an arbitrary (also time-independent) observable $\hat{A}$. We thus have $\delta_\mathcal{E} \delta_A \geq \frac{1}{2}|\langle [\hat{A}, \hat{H}] \rangle|$. Now, according to Ehrenfest's theorem,[3,18,19] $d\langle A \rangle/dt = -(i/\hbar)\langle [\hat{A}, \hat{H}] \rangle$. Therefore, $\delta_\mathcal{E} \delta_A \geq \frac{1}{2}\hbar |d\langle A \rangle/dt|$. Suppose now that the uncertainty $\delta_t$ associated with time is defined such that $\delta_A/\delta_t = |d\langle A \rangle/dt|$; in other words, the time-rate-of-change of the expected value $\langle A \rangle$ of the observable $\hat{A}$ is such that, during the time interval $\delta_t$, the expected value of $\hat{A}$ moves by one standard deviation $\delta_A$, thus rendering the motion perceptible. We will then have $\delta_\mathcal{E} \delta_t \geq \frac{1}{2}\hbar$.

**Acknowledgement**. The author is grateful to Prof. Jeff Lundeen of the University of Ottawa and Prof. Ewan Wright of the University of Arizona for many illuminating discussions.

## References

1. N. Bohr, "Discussions with Einstein on Epistemological Problems in Atomic Physics," in *Albert Einstein: Philosopher-Scientist* (edited by P. A. Schilpp), Library of Living Philosophers, Evanston, pp 200-241 (1949); reprinted in *Quantum Theory and Measurement* (editors J. A. Wheeler and W. H. Zurek), Princeton University Press, New Jersey (1983).
2. M. O. Scully, B-G. Englert, and H. Walther, "Quantum optical tests of complementarity," *Nature* **351**, pp 111-16 (May 9, 1991). https://doi.org/10.1038/351111a0.
3. R. P. Feynman, R. B. Leighton, and M. Sands, *The Feynman Lectures on Physics* (Volume III), Addison-Wesley, Massachusetts (1965).
4. G. Grynberg, A. Aspect, and C. Fabre, *Introduction to Quantum Optics*, Cambridge University Press, Cambridge, United Kingdom (2010).
5. M. Mansuripur, *Field, Force, Energy and Momentum in Classical Electrodynamics* (revised edition), Bentham Science Publishers, Sharjah (2017).
6. M. Born and E. Wolf, *Principles of Optics* (7$^{th}$ edition), Cambridge University Press, United Kingdom (1999).
7. L. Mandel and E. Wolf, *Optical Coherence and Quantum Optics*, Cambridge University Press, United Kingdom (1995).
8. G. Sagnac, *Comptes Rendus de l'Académie des Sciences* (*Paris*) **157**, 708-710, 1410-1413 (1913).
9. E. J. Post, "Sagnac effect," *Rev. Mod. Phys.* **39**, 475-493 (1967).
10. M. Mansuripur, *Classical Optics and its Applications* (2$^{nd}$ edition), Cambridge University Press, United Kingdom (2009).
11. J. D. Jackson, *Classical Electrodynamics* (3$^{rd}$ edition), Wiley, New York (1999).
12. R. Mir, J. S. Lundeen, M. W. Mitchell, A. M. Steinberg, J. L. Garretson, and H. M. Wiseman, "A double-slit 'which way' experiment on the complementarity-uncertainty debate," *New Journal of Physics* **9**, 287-97 (2007).
13. M. Mansuripur, "Insights into the behavior of certain optical systems gleaned from Feynman's approach to quantum electrodynamics," *Proceedings of SPIE* (this conference).
14. M. O. Scully and K. Drühl, "Quantum eraser: A proposed photon correlation experiment concerning observation and 'delayed choice' in quantum mechanics," *Physical Review A* **25**, 2208-13 (1982).
15. P. Storey, S. Tan, M. Collett, and D. Walls, "Path detection and the uncertainty principle," *Nature* **367**, 626-628 (1994).
16. B-G. Englert, M. O. Scully, and H. Walther, "Complementarity and uncertainty," *Nature* **375**, 367-368 (1995).
17. U. Eichmann, J. C. Bergquist, J. J. Bollinger, J. M. Gilligan, W. M. Itano, D. J. Wineland, and M. G. Raizen, "Young's Interference Experiment with Light Scattered from Two Atoms," *Phys. Rev. Lett.* **70**, 2359-62 (1993).
18. S. Gasiorowics, *Quantum Physics*, Wiley, New York (1974).
19. A. Yariv, *An Introduction to Theory and Applications of Quantum Mechanics*, Wiley, New York (1982).